\documentclass[aip,jap,superscriptaddress,amsmath,amssymb,reprint]{revtex4-1}
\usepackage{graphicx}
\usepackage{bm}

\begin{document}

\title{Study of negative thermal expansion in the frustrated spinel ZnCr$_{2}$Se$_{4}$}
\author{X. L. Chen}
\author{Z. R. Yang}
\email[Corresponding author: ]{zryang@issp.ac.cn}
\affiliation{Key Laboratory of Materials Physics, Institute of Solid State Physics, Chinese Academy of Sciences, Hefei 230031, People's Republic of China}
\author{W. Tong}
\author{Z. H. Huang}
\author{L. Zhang}
\author{S. L. Zhang}
\affiliation{High Magnetic Field Laboratory, Chinese Academy of Sciences, Hefei 230031, People's Republic of China}
\author{W. H. Song}
\affiliation{Key Laboratory of Materials Physics, Institute of Solid State Physics, Chinese Academy of Sciences, Hefei 230031, People's Republic of China}
\author{L. Pi}
\affiliation{High Magnetic Field Laboratory, Chinese Academy of Sciences, Hefei 230031, People's Republic of China}
\affiliation{University of Science and Technology of China, Hefei 230026, People's Republic of China}
\author{Y. P. Sun}
\affiliation{Key Laboratory of Materials Physics, Institute of Solid State Physics, Chinese Academy of Sciences, Hefei 230031, People's Republic of China}
\affiliation{High Magnetic Field Laboratory, Chinese Academy of Sciences, Hefei 230031, People's Republic of China}
\author{M. L. Tian}
\affiliation{High Magnetic Field Laboratory, Chinese Academy of Sciences, Hefei 230031, People's Republic of China}
\author{Y. H. Zhang}
\affiliation{High Magnetic Field Laboratory, Chinese Academy of Sciences, Hefei 230031, People's Republic of China}
\affiliation{University of Science and Technology of China, Hefei 230026, People's Republic of China}

\date{\today}

%\pacs{75.30.Et,75.50.Ee,78.30.-j}{Exchange and superexchange interactions}
%\pacs{75.50.Ee}{Antiferromagnetics}
%\pacs{78.30.-j}{Infrared and Raman spectra}

\begin{abstract}
The origin of negative thermal expansion (NTE) in the bond frustrated ZnCr$_{2}$Se$_{4}$ has been explored. ESR and FTIR document an ideal paramagnetic state above 100 K, below which ferromagnetic clusters coexist with the paramagnetic state down to $T_\mathrm{N}$. By fitting the inverse susceptibility above 100 K using a modified paramagnetic Curie-Weiss law, an exponentially changeable exchange integral $J$ is deduced. In the case of the variable $J$, magnetic exchange and lattice elastic energy couple with each other effectively via magnetoelastic interaction in the ferromagnetic clusters, where NTE occurs at a loss of exchange energy while a gain of lattice elastic one.
\end{abstract}

\maketitle

\section{Introduction}
Magnetic frustrated systems have recently been a subject of continuing interests for a manifold of fascinating states such as spin ice, spin liquid and orbital glass \textit{et al.} may be surviving down to $T$ = 0.\cite{Gardner,Balents,Tong} Among the reported materials, chromium-based spinels with the formula ACr$_{2}$X$_{4}$ (X = O, S, Se) are of essential role not only of being theoretical interests but also in exploring potential multi-functional materials.\cite{Baltzer,Yan,Hemberger,Kim} For instance, in CdCr$_{2}$S$_{4}$, the ferroelectricity and colossal magnetocapacitive coupling were observed.\cite{Hemberger-nature} The compound was suggested to be a multiferroic relaxor. However, Scott and coworkers argued its correctness and rather related it to be a conductive artefact.\cite{Catalan} Recently, the conclusion of conductive artefact is further evidenced by Yang \textsl{et al}., both experimentally and theoretically.\cite{Yang-JAP,Yang-EPL} As a result, the design or tuning of multi-functional materials for potential applications enables the necessity to clear what stands behind the novel phenomena observed.

ZnCr$_{2}$Se$_{4}$ with negative thermal expansion (NTE) is a case in point.\cite{Hemberger,Yokaichiya} This compound exhibits a large positive Curie-Weiss temperature.\cite{Lotgering} However, neutron diffraction study revealed at $T_\mathrm{N} = 21$ K a complex helical spins consisted of FM layers along [001] with a turning angle of $42{}^{\circ}$ between the adjacent ones.\cite{Plumier,Akimitsu,Hidaka} Interestingly, this spin configuration allows a field-induced electric polarization revealing magnetoelectric effect or multiferroicity.\cite{Siratori-ME,Murakawa} While early X-ray diffraction (XRD) and neutron powder diffraction indicated a tetragonal structural transformation at $T_\mathrm{N}$, subsequent neutron and synchrotron radiation results on single crystal showed an orthorhombic phase.\cite{Akimitsu,Hidaka,KK-XRD} Moreover, recent neutron powder diffraction found no sign of structural transition.\cite{Yokaichiya} In fact, the exact lattice symmetry is still subject of debate. On the other hand, IR spectroscopy experiment reported a marked splitting of the low-frequency mode below $T_\mathrm{N}$.\cite{Rudolf} This elucidates an essential spin-phonon coupling. Both geometrical and additional bond frustration are believed to be crucial factors.\cite{Lee,Hemberger} In addition, since Cr$^{3+}$ in an octahedral crystal field is Jahn-Teller inactive, spin-phonon coupling is of vital importance in lifting frustration. Nevertheless, the very origin of NTE is yet far from being well understood.

In the present paper, a set of experimental techniques is utilized to probe the spin-lattice correlation in ZnCr$_{2}$Se$_{4}$. By considering a variable exchange integral $J$ with respect to temperature or lattice constant, NTE is attributed to a result of the competition between magnetic exchange and lattice elastic energy via magnetoelastic coupling.

\section{Experiments}
The polycrystalline sample of ZnCr$_{2}$Se$_{4}$ was prepared by standard solid state reaction method. High purity powders of zinc (99.9\%), chromium (99.9\%) and selenium (99.9\%) were mixed according to the stoichiometric ratio. Next, the powders were sealed in an evacuated quartz tube, and heated slowly to 850 ${}^{\circ}\mathrm{C}$ in seven days. Then the sample was reground, pelletized, sealed and heated again for another three days at 850 ${}^{\circ}\mathrm{C}$. The magnetic data was collected on a Quantum Design superconducting quantum interference device (SQUID) magnetometer. The temperature dependent XRD patterns were obtained using XRD (Rigaku TTRIII). ESR measurements were performed using a Bruker EMX plus 10/12 CW-spectrometer at X-band frequencies ($\upsilon$ = 9.39 GHz) in a continuous He gas-flow cryostat for 2--300 K. The transmittance spectra were collected in the far-infrared range using the Bruker Fourier-transform spectrometer Vertex 80v equipped with a He bath for 5--300 K.

\section{Results and discussion}
\begin{figure}[htb]
\centering
\includegraphics[scale=1]{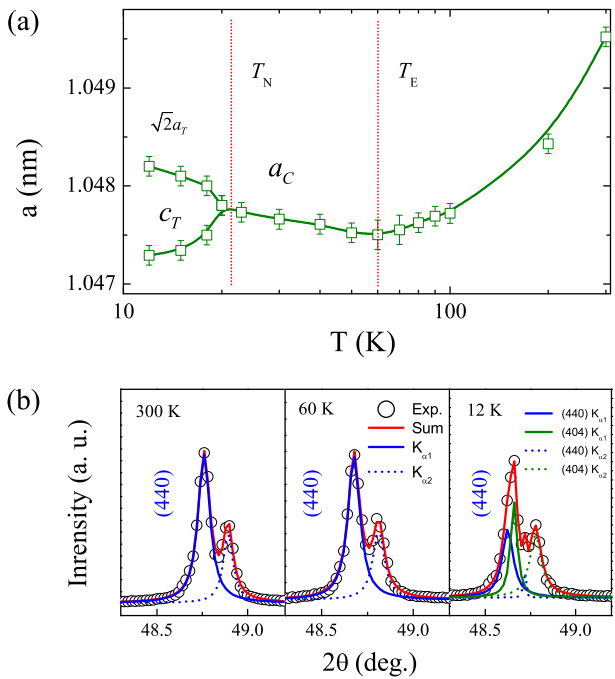}
\caption
{
(Color online) (a) Temperature dependent lattice parameter $a$ vs $T$ in semilogarithmic for ZnCr$_{2}$Se$_{4}$. Error bars are average of repeated fittings. The negative thermal expansion initial temperature $T_\mathrm{E}$ and the antiferromagnetic order one $T_\mathrm{N}$ are drawn in red doted lines. (b) The representative peaks (440) at 300 K, 60 K and 12 K. Circles are experimental data; Solid and dashed lines are Lorentzian fits.
}
\label{XRD}
\end{figure}

The XRD data was analyzed using the standard Rietveld technique, which shows a single-phase material with cubic spinel structure at room temperature. Figure~\ref{XRD}(a) shows the temperature ($T$) dependence of lattice constant $a$. With decreasing temperature, $a$ first decreases rapidly and then manifests a negative thermal expansion behavior below about $T_\mathrm{E} = 60$ K, which is almost concordant with the previous results.\cite{Hemberger,Yokaichiya} Upon further cooling below 20 K, splitting of several peaks in XRD spectra are observed. The representative peak (440) at 12 K that splits into (440) and (404) is presented in Fig.~\ref{XRD}(b). This signals a cubic to tetragonal structural transition with space group \textsl{I4$_{1}$/amd}.\cite{Akimitsu,KK-XRD}
 
\begin{figure}[htb]
\centering
\includegraphics[scale=1]{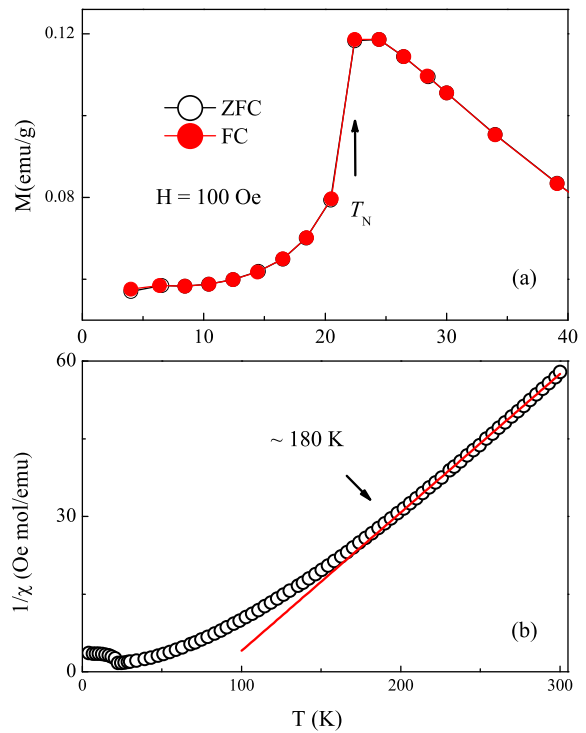}
\caption
{
(Color online) (a) Low temperature dependence of the magnetization $M$ for a polycrystalline sample ZnCr$_{2}$Se$_{4}$ at 100 Oe. (b) A Curie-Weiss fitting of the inverse susceptibility.
}
\label{MT}
\end{figure}

Figure~\ref{MT}(a) presents the magnetization ($M$) versus $T$ at low applied magnetic field of 100 Oe under both zero-field-cooled (ZFC) and field-cooled (FC) sequences. At about $T_\mathrm{N} = 22$ K, $M$ shows a sharp AFM transition.\cite{Hemberger,Plumier,Akimitsu,Hidaka,Yokaichiya} Figure~\ref{MT}(b) shows a fitting of the inverse susceptibility 1/$\chi$ according to the paramagnetic (PM) Curie-Weiss law \(\frac{1}{\chi}=\frac{T-\Theta_\mathrm{CW0}}{C}\). A large positive Curie-Weiss temperature $\Theta _\mathrm{CW0}$ = 85 K and the coefficient $C = 3.74$ are obtained. The effective magnetic moment calculated by the formula $\mu _\mathrm{eff} = 2.83\sqrt\mathrm{C/2}$ (in CGS units) equals to $3.87\mu _\mathrm{B}$, in agreement with the spin-only Cr$^{3+}$ ion.

\begin{figure}[htb]
\centering
\includegraphics[scale=1]{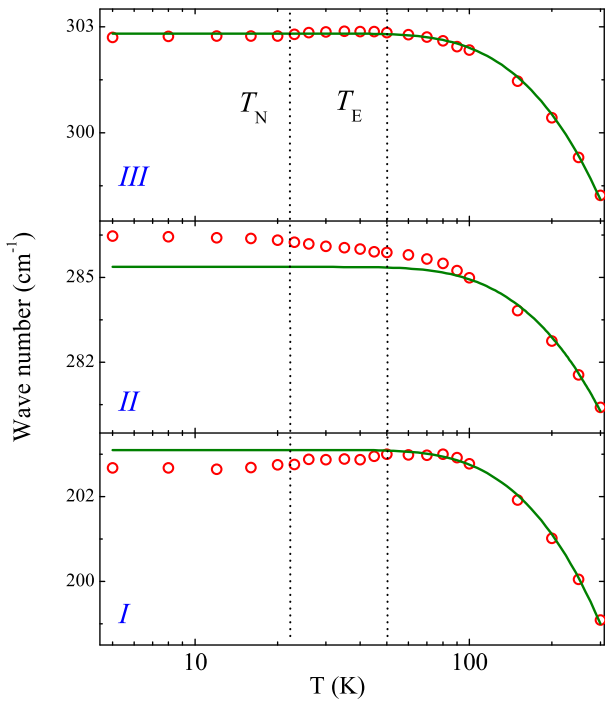}
\caption
{
(Color online) Temperature dependence of wave number of the infrared activated mode $I$, $II$ and $III$ for ZnCr$_{2}$Se$_{4}$. The solid curve is a fit described in the text. The characteristic temperatures are indicated.
}
\label{FTIRF}
\end{figure}

The large positive Curie-Weiss temperature implies a dominant FM exchange interaction, however, the compound shows an AFM ordering at low temperatures. Hence, we first investigate the lattice dynamic inspired by the strong spin-lattice correlation. Within the wave-number range inspected, three infrared modes, labeled as $I$, $II$ and $III$ are observed simultaneously. Generally speaking, the phonon eigenfrequency follows an anharmonic behavior, which can be described by the following formula:

\begin{equation}\label{eq1}
\omega_{i} = \omega_{0i}\left[1 - \frac{\alpha_{i}}{\mathrm{exp}(\Theta/T)-1}\right],
\end{equation}

where $\omega_{0i}$, $\alpha_{i}$ and $\Theta$ are the eigenfrequency of mode $i$ with decoupling of the spin and phonon at 0 K, the weight factor of mode $i$ and the Debye temperature $\Theta$ = 309 K.\cite{Rudolf} A detailed temperature dependence of eigenfrequency is exhibited in Fig.~\ref{FTIRF}. The solid line is a fitting of classical anharmonic behavior according to equation \ref{eq1}. Mode $I$ shows negative shifts below $T_\mathrm{E}$ and Mode $II$ shows positive shifts from about 100 K compared to the normal anharmonic behaviors, respectively. According to Lutz \textsl{et al.} and coworkers, the highest frequency mode III is ascribed to the Cr-Se vibration, however, modes II and I originate from combined vibrations of Cr-Se and Zn-Se. For mode II, the ratio of contribution from Cr-Se and Zn-Se is 78 : 18, while for mode I the ratio of Cr-Se to Zn-Se is 21 : 73.\cite{Zwinscher} Cr-Se bond relates to the ferromagnetic (FM) spin super-exchange interaction and Zn-Se links to the antiferromagnetic (AFM) super-exchanges.\cite{Baltzer} As a result, FM Cr-X-Cr bonds dominate modes II and III while AFM linkages Cr-X-A-X-Cr determine mainly the eigenfrequency of mode I.\cite{Rudolf-NJP,Wakamura} The deviation of mode II below 100 K due to spin-phonon coupling indicates that FM fluctuations involve already at this temperature. Furthermore, the opposite shifts of modes I and II in the NTE temperature region may refer to a dynamic competition of FM and AFM superexchanges.

\begin{figure}[htb]
\centering
\includegraphics[scale=1]{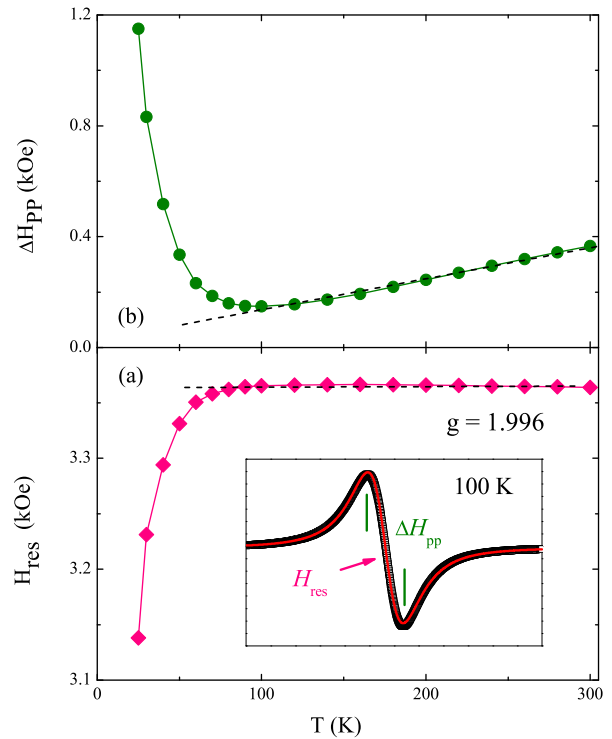}
\caption
{
(Color online) (a) Temperature dependence of the resonance field $H_\mathrm{res}$. Inset: a typical fitting of the ESR spectrum at 100 K, where the definitions of the parameters are shown. (b) The peak-to-peak linewidth $\Delta H_\mathrm{PP}$ vs $T$. The dashed lines are guided by eyes.
}
\label{ESR}
\end{figure}

The temperature dependent ESR spectrum is further investigated. A PM signal is observed at room temperature and it vanishes below $T_\mathrm{N}$ (not drawn). The resonant field ($H_\mathrm{res}$) and peak-to-peak line-width ($\Delta H_\mathrm{PP}$) as a function of temperature are plotted in Fig.~\ref{ESR}(a) and (b), respectively. A typical ESR spectrum at 100 K and fitting using the first-order derivative symmetric Lorentzian function are presented in the inset of Fig.~\ref{ESR}(a). As can be seen from Fig.~\ref{ESR}(a), with decreasing temperature from 300 K to 100 K, $H_\mathrm{res}$ shows a constant value while $\Delta H_\mathrm{PP}$ decreases almost linearly. Upon further cooling, $H_\mathrm{res}$ drops drastically and accordingly $\Delta H_\mathrm{PP}$ broadens greatly, in accordance with Ref. \cite{Hemberger}.

The g factor is 1.996, in agreement well with the previous works.\cite{Hemberger,Stickler,Siratori} The linear decrease of $\Delta H_\mathrm{PP}$ with temperature can be attributed to a single-phonon spin-lattice relaxation mechanism.\cite{Rettori,Huber} Both the constant g-factor and linear behavior of $\Delta H_\mathrm{PP}$ evidence a well-defined paramagnetic (PM) state at least above 100 K. Note that $H_\mathrm{res}$ starts to decrease and $\Delta H_\mathrm{PP}$ broadens at 100 K. Meanwhile, an FM-related positive shift for Mode II is observed in the IR modes in Fig.~\ref{FTIRF}. These can thereby be correlated to the onset of FM fluctuations and spin-phonon coupling. As we know, $H_\mathrm{res}$ is a sum of $H_\mathrm{int}$ and $H_\mathrm{ext}$, where $H_\mathrm{int}$ and $H_\mathrm{ext}$ are the equivalent internal and the external actural fields, respectively. For the case $H_\mathrm{int}$ $>$ 0, the internal field may shift the resonance line to lower field; on the contrary, the resonance signal would appear at higher field with negative internal magnetic field. The decreasing of $H_\mathrm{res}$ below 100 K indicates an increasing of $H_\mathrm{int}$, which may be caused by the interaction between the localized magnetic moments and the demagnetization effect.\cite{Rettori} Therefore, the enhancement of $H_\mathrm{int}$ and onset of FM fluctuations reveal gradually growing FM clusters, forming an FM-cluster and PM mixed state from 100 K to $T_\mathrm{N}$. The existence of FM clusters is also supported by the following NTE analysis. Below $T_\mathrm{N}$, the signal disappears due to the AFM ordering transition.

As is discussed above, the system keeps a pure PM state at least above 100 K, so the inverse susceptibility should be described by the PM Curie-Weiss law down to this temperature. However, it departs from the linear behavior at a temperature as high as about 180 K, see Fig.~\ref{MT}(b). Recalling the fitting process in Fig.~\ref{MT}(b), we have assumed a constant Curie-Weiss temperature $\Theta_\mathrm{CW0}$, i.e., a constant magnetic exchange interaction $J$. The behavior of IR modes below 100 K implies a competition of FM and AFM superexchange interactions. In addition, the nearest neighbor FM Cr-Se-Cr and other neighbor AFM Cr-Se-Zn-Se-Cr superexchange interactions depend strongly on the lattice constant.\cite{Baltzer} It means that the total $J$ may change since $a$ decreases dramatically upon cooling [Fig.~\ref{XRD}(a)]. Accordingly, the traditional Curie-Weiss behavior should be modified within the present case to bridge the gap mentioned above. In specific, one should take a variable $\Theta _\mathrm{CW}$ (or $J$) as a function of $T$ or $a$ into account.

\begin{figure}[htb]
\centering
\includegraphics[scale=1]{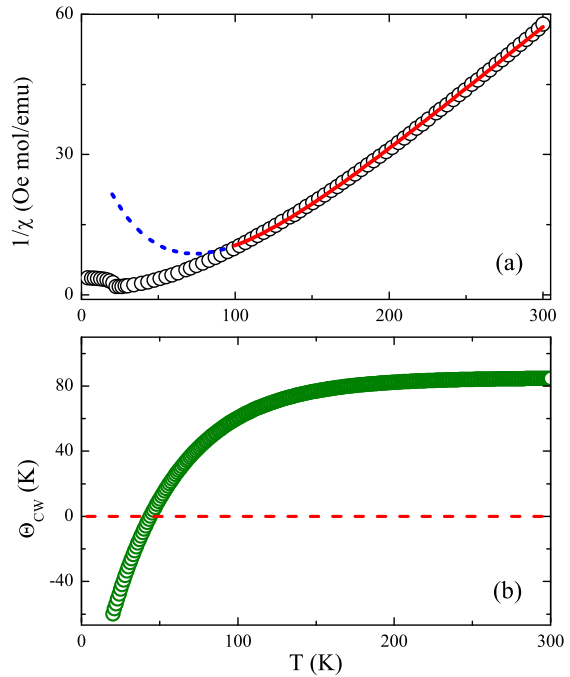}
\caption
{
(Color online) (a) A fitting of the inverse susceptibility above 100 K in red solid line using \(\frac{1}{\chi}=\frac{T-\Theta_\mathrm{CW}(T)}{C}\) taking a temperature or lattice constant $a$ dependent Curie-Weiss temperature into account. The short dashed blue line is an extension of the fitting to low temperatures. (b) $\Theta_\mathrm{CW}$ vs $T$ and the red dashed line indicates $\Theta_\mathrm{CW}$ = 0 at 45 K.
}
\label{Tcw}
\end{figure}

In AFM spinel oxides, the Curie-Weiss temperature changes exponentially with the lattice parameter.\cite{Rudolf-NJP} Naturally, an empirical description of $\Theta_\mathrm{CW}(T)$ = $\Theta_\mathrm{CW0} - \alpha \times e^{-T/\beta}$ is postulated. The fitting of the inverse susceptibility above 100 K using the modified Curie-Weiss behavior \(\frac{1}{\chi}=\frac{T-\Theta_\mathrm{CW}(T)}{C}\) is exhibited in Fig.~\ref{Tcw}(a). The parameters are $\alpha$ = 226 and $\beta$ = 45. Furthermore, a remarkable deviation below 100 K in blue short dashed line indicates the appearance of the effective internal field originating from FM clusters. Next, based on the obtained $\alpha$ and $\beta$, $\Theta_\mathrm{CW}(T)$ is extrapolated to low temperatures as exhibited in Fig.~\ref{Tcw}(b). It shows a derivation at about 180 K from the nearly constant value. With further lowering temperature, it decreases faster and faster and below $T$ $\approx$ 45 K, $\Theta_\mathrm{CW}(T)$ even becomes negative. These features may interpret qualitatively the fact that ZnCr$_{2}$Se$_{4}$ is dominated by ferromagnetic exchange interaction but orders antiferromagnetically at low temperatures. Since the exchange integral and the Curie-Weiss temperature are linked by $J(T) \varpropto \Theta_\mathrm{CW}(T)$ [$J(a) \varpropto \Theta_\mathrm{CW}(a)$], we will use $J(T)$ [$J(a)$] instead in the following discussion.

Given that $J$ is changeable, magnetic exchange and lattice elastic energies can link effectively with each other via magnetoelastic coupling. The free energy $F$ in a magnetoelastic system is expressed as

\begin{equation}
F(T) = -J(T) \sum_{i,j} \overrightarrow{S_{i}} \cdot \overrightarrow{S_{j}}
+ \frac{1}{2} N \omega ^{2} \triangle ^{2}(T) -T \cdot S(T),
\end{equation}

where N the number of the ion sites, $\omega$ the averaged vibrational angular frequency and $\triangle$ the averaged strain relative to the equilibrium lattice constant. The first term is exchange energy ($E_{ex}$) as a function of $J$, the second is lattice elastic energy ($E_{el}$) related mainly to the lattice parameter $a$ (or equivalent $T$) and the last is the entropy. From the above equation we know that if the system stays at an ideal PM state, then $E_{ex}$ = 0. So $E_{ex}$ and $E_{el}$ will show no coupling, as is observed from the IR spectra above 100 K.

When the system stays at an FM state with a changeable $J$, there may exist a competition between $E_{ex}$ and $E_{el}$ since the former is negative while the latter always positive. Indeed, it has been concluded above that $J$ decreases exponentially [Fig.~\ref{Tcw}(a)] and some FM clusters forms gradually below 100 K. Therefore, the concomitant decrease of $J$ and $a$ causes an increasing $E_{ex}$ but decreasing $E_{el}$ in the FM clusters. At some critical point, a totally compensation between them may present. If $a$ further decreases as cooling, the variation of $E_{ex}$ would gradually exceed that of $E_{el}$ in magnitude. Especially when $J$ drops sharply with respect to $a$, say here at $T_\mathrm{E}$, a tiny decrement of $a$ will give rise to a dramatic increment of $E_{ex}$ whereas $E_{el}$ keeps nearly constant. The state in a system subjected to stimuli, such as cooling, always tends to develop towards one that can lower $F$. In this sense it is favorable to lowering $F$ by expanding the lattice parameter $a$ to increase $J$ ($J > 0$) and thereby to decrease $E_{ex}$ due to its negative value, at the same time at a cost of few increases of $E_{el}$ in magnitude. This means that a negative thermal expansion of the lattice originating from the FM clusters with a exponentially changeable $J$ is expected. It is worthy of noting that when applying a magnetic field to the system in the NTE temperature region, NTE in magnitude enhances.\cite{Hemberger} This is because the size or population of the FM clusters increases when applying a magnetic field.

On the contrary, when an AFM ordering appears, $J$ becomes negative and the condition to stimulate NTE is not met any more. Normal thermal expansion upon cooling results in a simultaneous decreasing of $E_{ex}$ and $E_{el}$, which is consistent to lowering $F$. In fact, a normal expansion feature is observed below $T_\mathrm{N}$.\cite{Hemberger} It should be noted that the existence of NTE \cite{Hemberger,Yokaichiya} evidences in turn that $J$ is changeable. If $J$ keeps constant in a FM cluster, $E_{ex}$ will be almost constant and a normal thermal expansion of the lattice alone can give rise to a decrease of $E_{el}$ and of $F$ sufficiently.

\section{Conclusions}
To summarize, we have investigated the origin of NTE in strongly bond frustrated ZnCr$_{2}$Se$_{4}$. By fitting the inverse susceptibility above 100 K, an exponentially variable exchange $J$ is deduced. The exchange and lattice elastic energy can effectively couple with each other via magnetoelastic on basis of this changeable $J$. NTE is qualitatively interpreted as a competition between the two kinds of energy.

\begin{acknowledgments}
This research was financially supported by the National Key Basic Research of China Grant, Nos. 2010CB923403, and 2011CBA00111, and the National Nature Science Foundation of China Grant 11074258.
\end{acknowledgments}

\end{document}